\documentclass[aps,prl,twocolumn,10pt,a4paper,showpacs,floatfix,footinbib]{revtex4-1}
\usepackage{amssymb}
\usepackage{graphicx}
\usepackage{epsfig}
\usepackage{amsmath}
\usepackage{slashed}
\usepackage{units}
\usepackage{hyperref}
\usepackage{array}
\usepackage{lmodern}
\usepackage{dcolumn}
\usepackage{bm}
\usepackage{tabularx}
\usepackage{inputenc}

\immediate\write18{bibtex \jobname}

\begin{document}

\title{Quantum vacuum photon modes and superhydrophobicity}

\author{Louis Dellieu$^1$}
\email{louis.dellieu@unamur.be}
\affiliation{Research Center in Physics of Matter and Radiation (PMR), Department of Physics, University of Namur, 61 rue de Bruxelles, B-5000 Namur, 
Belgium}

\author{Olivier Deparis}
\affiliation{Research Center in Physics of Matter and Radiation (PMR), Department of Physics, University of Namur, 61 rue de Bruxelles, B-5000 Namur, 
Belgium}

\author{J\'{e}r\^{o}me Muller}
\affiliation{Research Center in Physics of Matter and Radiation (PMR), Department of Physics, University of Namur, 61 rue de Bruxelles, B-5000 Namur, 
Belgium}

\author{Micha\"{e}l Sarrazin$^1$}
\email{michael.sarrazin@unamur.be}
\affiliation{Research Center in Physics of Matter and Radiation (PMR), Department of Physics, University of Namur, 61 rue de Bruxelles, B-5000 Namur, 
Belgium}

\begin{abstract}
Nanostructures are commonly used for developing superhydrophobic surfaces. 
However, available wetting theoretical models ignore the effect of vacuum photon-modes alteration 
on van der Waals forces and thus on hydrophobicity. Using first-principle calculations, 
we show that superhydrophibicity of nanostructured surfaces is
dramatically enhanced by vacuum photon-modes tuning. As a case study, wetting contact angles 
of a water droplet above a polyethylene nanostructured surface are obtained from the interaction potential energy 
calculated as function of the droplet-surface separation distance. This new approach could pave the way for the design
of novel superhydrophobic coatings.\\
$^1$\textit{These authors have contributed equally to this work.}
\end{abstract}

\maketitle
While superhydrophobicity on structured surfaces (Fig. 1a) is a topic of high interest \cite{1,2} and wetting phenomena are known to be related to van der Waals forces \cite{2b,3,4,5}, the influence of surface nano-corrugations on van der Waals forces has not been considered in wetting theoretical models as of yet \cite{5b,5c,5cd,5cde,5d,5e}.
In this letter, we address this point and shed a new light on an old yet interesting problem by revisiting the role of nano-structuration on wetting phenomena in the framework of the quantum electrodynamics description of van der Waals interactions.  

Casimir interactions \cite{6}, which generalize van der Waals interactions between structured surfaces, have been extensively studied in case of metals and semi-conductors \cite{7,8,9,10}. For this purpose, many authors use an extension of the Lifshitz theory \cite{11} of van der Waals interactions between macroscopic bodies \cite{10,11,12,13}. Considering media 1 and 2 occupying the half-spaces $z < 0$ and $z > L$, respectively, and separated by vacuum, it can be shown that the van der Waals interaction potential energy $U$ is given by \cite{10,11,12,13} $U=\sum_p\frac{1}{2} \hbar (\omega _{p}(L) - \omega _{p}(L \rightarrow \infty))$ where $\omega_{p}(L)$ is the eigen angular frequency - for a given polarization - of the p$^{th}$ vacuum photon-mode available between the two media facing each other. Indeed, the van der Waals force results from the exchange of virtual photons between both interacting bodies (Fig. 1b) \cite{14}. Using the Cauchy's argument principle of the complex analysis and considering the analytical properties of the Fresnel coefficients related to each body, the interaction energy can be expressed as \cite{10,12,13}:

\begin{figure}[b]
\centerline{\ \includegraphics[width=7 cm]{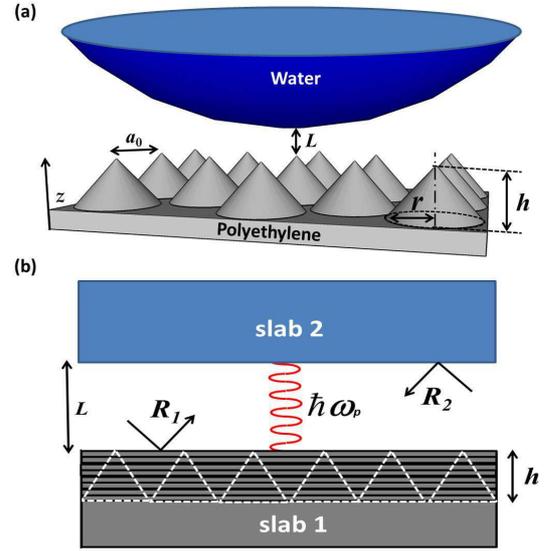}}
\caption{(Color online). (a) Corrugated surface separated by a distance $L$  from a
water droplet. The surface of the droplet can be considered as flat at the scale of corrugation. (b) Approximation of the two-body system. $R_1$ and $R_2$ are the reflection coefficients of the effective multilayer (slab 1) and of the water droplet (slab 2), respectively. Virtual photons $\hbar \omega_p$ exchanged between the two slabs are pictured in red.}
\label{fig1}
\end{figure}

\begin{eqnarray}
U(L) &=&\frac \hbar {2\pi }\sum_{m=s,p}\int \frac{d^2k_{//}}{(2\pi )^2}%
\int_0^\infty d\xi  \label{3} \\
&&\times \ln (1-R_1^m(i\xi ,\mathbf{k}_{//})R_2^m(i\xi ,\mathbf{k}%
_{//})e^{-2\kappa L})  \nonumber
\end{eqnarray}
where $\kappa =\sqrt{\frac{\xi ^2}{c^2}+\left| \mathbf{k}_{//}\right| ^2}$, $R_1$$^m$ ($R_2$$^m$) is the complex reflection coefficient of slab $1$
(slab $2$) (Fig. 1b) in the $m$ polarization state ($s$ or $p$ states) and $k_{//}$ is the parallel component of the photon wave vector. The use of the complex angular frequency $\omega$ $=$ $\textit{i}\xi$ arises from numerical computation considerations. In deriving Eq. (1), the dependence of mode free energy on temperature $T$ can be neglected since, around ambient temperature, $\hbar\omega_p$$>>$$k_BT$ is satisfied for all photon energies involved here.
It is noteworthy that, at short distances ($L \leq 10$ nm) \cite{14c}, Eq. (1) is well approximated \cite{14d} by the well-known Hamaker expression \cite{14b}:

\begin{equation}
U(L) = -\frac{A_h}{12\pi L^2} , \label{5} 
\end{equation}
where $A_h$ is the effective Hamaker constant which can be deduced from the computed energy, i.e. Eq. (1).

Moreover, it is known that the Hamaker theory, when applied to molecular solids, is able to predict the contact angle of a water droplet on a surface \cite{4}. Indeed, from the van der Waals potential energy calculated between a molecular solid and a liquid, we can deduce immediately the corresponding contact angle $\theta$ \cite{4} :

\begin{equation}
\cos (\theta )=-1+\frac{\left|U(L_0)\right|}{\gamma _l} ,  \label{4}
\end{equation}
where $U(L_0)$ is the potential energy between the two media separated by the distance $L_0$, which is the equilibrium
separation distance \cite{14c} between the water droplet and the surface.
This distance, originally defined for a flat surface \cite{14c}, is assumed to remain the same when the solid surface is corrugated. Indeed, the lack of data on the equilibrium separation distance values for corrugated surfaces leaves us no choice but to use values for flat surfaces of identical materials. However, this bold hypothesis does not affect the main results of our model \cite{14e}.
In Eq. (3), $\gamma _l$ is the liquid surface
tension, which for water is $72.5$ mN/m \cite{15}. Until now, previous studies used Eq. (3) in the context of flat interfaces only \cite{4}. The present work extends its use to corrugated surfaces.

Following the above considerations, we can expect that corrugations covering the surface of a molecular solid should affect its wettability in a new perspective. Indeed, in the same way as corrugations on a surface modify its reflectivity \cite{17,18}, their effects on the van der Waals interactions can no longer be ignored. According to Eq. (1), modifying the reflection coefficients should dramatically influence the interaction energy and therefore the wetting properties - \textit{via} Eq. (3) - as soon as the interaction between a liquid and corrugated molecular solid is concerned. This prediction is crucial, far from being obvious since usual wetting models of nanostructured surfaces ignore any alteration of the van der Waals force. For instance, wettability is commonly described, in the simplest way, by Cassies law in a thermodynamic approach involving surface tensions and Gibbs energy minimization \cite{5b}. Basically, such an approach does not consider explicitly the van der Waals interactions. Similarly, finite element methods relying on Navier-Stokes equations \cite{5c,5cd,5cde}, suffer from the same drawback. Even molecular dynamic simulations \cite{5d,5e} which consider explicitly van der Waals interactions, do not take into account their alteration in the presence of surface corrugations.

Let us now further develop the approach. The two-body system under study consists of 
a nanostructured surface and a droplet of water separated by a distance $L$ (Fig. 1a). 
Since the water droplet is much bigger than the surface corrugation features, it can be described by a slab (Fig. 1b). The surface is nanostructured with cones of
height \textit{h} arranged on a hexagonal lattice with a lattice parameter chosen to be $a_0= 10$ nm. The cone base radius is $r = 5$ nm and the cone height 
$h$ ranges from $10$ nm to $100$ nm in order to vary the
antireflection character of the surface.  Such a geometry is known to improve the antireflection behavior of surfaces \cite{17,18}. In the following, the
nanostructured surface is assumed to be made of polyethylene. The choice
of polyethylene was motivated by the need to work with a molecular solid. Since the lattice parameter $a_0$ has
subwavelength dimensions (for wavelengths below $10$ nm, polyethylene permittivity is close to $1$), the corrugated surface can be described by a
continuous effective material with a graded refractive index along its thickness \cite{18} (Fig.1 b).
The effective dielectric function of the corrugated surface can be expressed by \cite{17,18}:

\begin{equation}
\varepsilon (z)=1 +(\varepsilon _{material}- 1)f(z)  \label{6}
\end{equation}
where $\varepsilon_{material}$ is the dielectric function of the bulk material and $f(z)$ is a filling factor given by: $f(z)=\pi r^2(z)/S$
with $S = a_0^2\sqrt{3}/2$ and $r(z)$ the radius of the
circular section of the cones at coordinate $z$. The system is then
reduced to a water slab and an effective multilayer (Fig. 1b).
In computations, polyethylene permittivity is described by a modified Lorentz oscillator
model \cite{23}:

\begin{equation}
\varepsilon (\omega )=\varepsilon _\infty +\sum_{p=1}^N\frac{\Delta
\varepsilon _p(\omega _p^2-i\gamma _p^{^{\prime }}\omega )}{\omega
_p^2-2i\omega \gamma _p-\omega ^2}  \label{8}
\end{equation}
where $\varepsilon _\infty$ is the permittivity at infinite frequency, $\omega _p$ is
the plasma frequency, $\gamma _p$ and $\gamma _p^{^{\prime }}$ are related to
relaxation processes associated to the\textit{\ p}$^{th}$
oscillator and $\Delta \varepsilon _p$ is defined such as $\sum_{p=1}^N$$\Delta \varepsilon _p$ = $\epsilon_{stat}$ - $\epsilon_\infty$ with $\epsilon_{stat}$ the permittivity at zero frequency. 
The values of these parameters are listed in Table 1 and were obtained by
fitting (\textit{N} = 2 oscillators) the experimental dielectric function of polyethylene which tends
to unity for wavelengths shorter than $10$ nm \cite{24}. The water slab is also described by a dielectric function whose analytical form and parameters were taken from \cite{25}. 

Actually, the water droplet is in equilibrium with water vapor. Therefore, the medium separating the surface and the droplet should be vapor instead of vacuum. However, since the water vapor dielectric constant is very close to that of vacuum at all wavelengths of interest and whatever the vapor partial pressure is \cite{25b}, there is no difference in considering a vacuum interface instead of a vapor interface from the point of view of electrodynamical calculations. Moreover, the water slab is considered to be in a Cassie state which assumes a flat meniscus \cite{5b}. In fact, a more realistic meniscus profile, like overhanging profile, could be modelled as a thin effective layer. This optically thin layer, however, would not lead to significant effects in the electrodynamical calculations of the van der Waals forces.

\begin{table}[t]
\begin{center}
\begin{tabular}{|l|l|l|}
\hline
 & $p=1$ & $p=2$ \\ 
\hline
$\Delta \varepsilon _p$ & $0.2479$ & $0.970$ \\ 
$\omega _p$ & $1.27 \times 10^{16}$rad s$^{-1}$ & $1.88 \times 10^{16}$rad s$^{-1}$ \\ 
$\gamma _p$ & $9.66 \times 10^{14}$rad s$^{-1}$ & $5.27 \times 10^{15}$rad s$^{-1}$ \\ 
$\gamma _p^{^{\prime }}$ & $1.26 \times 10^{16}$rad s$^{-1}$ & $3.63 \times 10^{15}$rad s$^{-1}$\\
\hline
\end{tabular}
\end{center}
\caption{Parameters of the dielectric function of polyethylene. }
\label{StorageExp}
\end{table}

\begin{figure}[t]
\centerline{\ \includegraphics[width=8 cm]{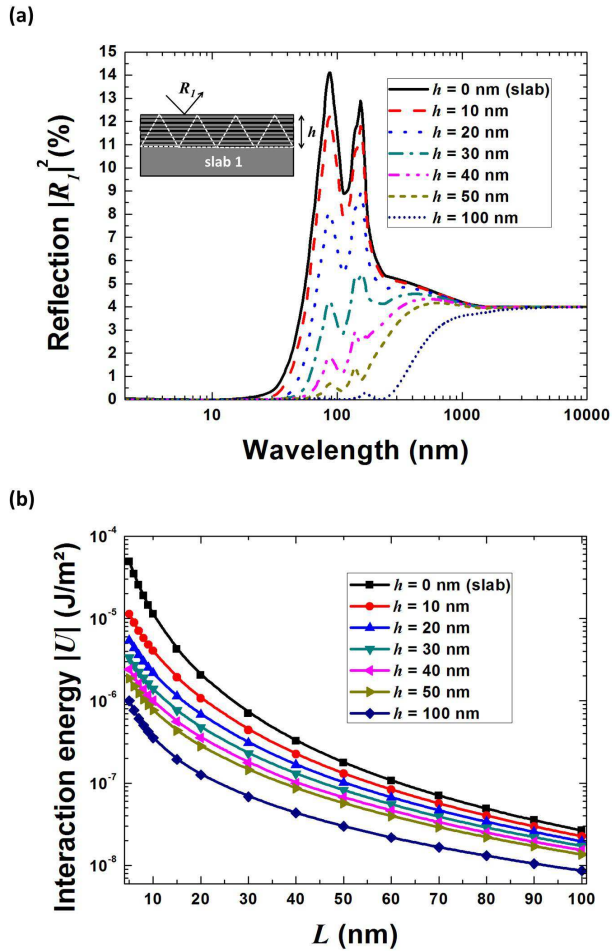}}
\caption{(Color online). (a) Intensity reflection coefficient (at normal incidence) of corrugated polyethylene slabs 
with cones of various heights $h$. (b) Interaction energy between a water droplet
and corrugated polyethylene slabs.}
\label{fig3}
\end{figure}

Thanks to a standard multilayer computational method \cite{26}, the intensity (square modulus) reflection coefficient of the nanostructured polyethylene surface $\left|R_1(\lambda)\right|^2$ was computed (see Fig. 2a where only normal incidence $k_{//}$ = 0 is considered for the sake of clarity).
Broadband antireflection character of the
corrugated surface was obtained by increasing the height of the cones: the higher the cones, the lower the reflection.
Moreover, by numerically solving Eq. (1) for each height, the interaction energy between the nanostructured surface and the water droplet was determined (Fig. 2b). The black line corresponds to the interaction energy between a flat polyethylene surface and a water droplet. As the height of cones is increased, the interaction energy is clearly altered: the higher the cones, the lower the potential (Fig. 2b). In other words, the van der Waals potential energy decreases as the antireflection character of the polyethylene surface increases (Fig. 2a). This fundamental point is the main finding of our study: the alteration of van der Waals potential energy due to the presence of surface corrugation, even in a Cassie state. This result can be roughly interpreted in a simple way. Indeed, when the cone height \textit{h} increases, the reflection coefficient $\left|R_1\right|^2$ decreases (Fig. 2a). As a result, the quality factor \textit{Q} of the Fabry-Perot cavity formed by the gap between the two slabs also decreases \cite{photonics}. Therefore, the electromagnetic energy stored in the Fabry-Perot cavity decreases \cite{photonics} i.e. the vacuum photon-modes contributions to the potential energy \textit{U} diminish. This explaining why the van der Waals interaction energy decreases as the nanostructure is tuned towards higher $h$ values (Fig. 2b).

The equilibrium separation distance between water and flat polyethylene surface was calculated to be $L_0 = 0.145$ nm \cite{4}. Although this value is reported for two flat polyethylene surfaces facing each other, it remains essentially unchanged while considering the present polyethylene-water system, as shown in \cite{4}. Due to numerical considerations, potential energies at $L = L_0$ are extrapolated from Eq. (2), by fitting the Hamaker constant $A_h$ to the calculated energy values for $L \leq 10$ nm \cite{14c}. The Hamaker constant for each height $h$ is shown in Fig. 3a. Using Eq. (2) and Eq. (3) with $L = L_0$, we can check the expected value of contact angle of a water droplet on a flat (\textit{h} = 0 nm) polyethylene surface: $\theta=102^\circ$ \cite{27} (Fig. 3b). 
For a corrugated surface, the contact angle and the Hamaker constant of the system are therefore modified with respect to a flat surface: the Hamaker constant decreases (Fig. 3a) while the contact angle dramatically increases (Fig. 3b) as the cone height increases. Superhydrophobicity ($\theta \geq 150^\circ$) is achieved for $h > 20$ nm here. The tuning of the optical properties of the polyethylene surface \textit{via} its nanostructuration directly affects its wettability. This finding is of practical importance as it will be discussed hereafter.

\begin{figure}[t]
\centerline{\ \includegraphics[width=8 cm]{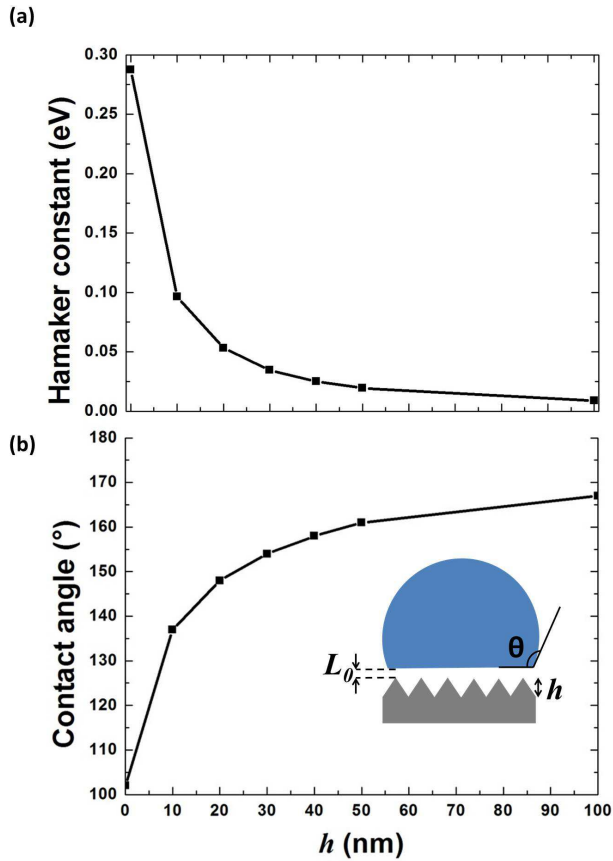}}
\caption{(Color online). (a) Evolution of the Hamaker constant as function of the height $h$ of cones.  (b) Evolution of the contact
angle of a water droplet on nanostructured polyethylene as function of the height $h$ of cones.}
\label{fig4}
\end{figure}

Our calculation predicts a contact angle of $173^{\circ }$
for $h=1\mu$m and of $175^{\circ }$ for $h=10 \mu$m (not shown on Fig. 3b). Ultimately, we
can expect a contact angle approaching $180^{\circ}$ for a very high aspect ratio of cones. According to Eq. 3, a contact angle  of $180^{\circ}$
corresponds to zero potential energy. The asymptotic increase of the contact angle towards $180^{\circ}$
as $h$ increases results from the fact that, beyond a height of $100$ nm (maximum value displayed on Fig. 3b), the potential energy $U$ decreases 
more slowly with $h$. Indeed, the reflection coefficient $\left|R_1\right|^2$ barely decreases at wavelengths $\lambda_p > 1000$ nm when $h$ increases (Fig. 2a). In addition, the related vacuum photon-modes angular frequencies $\omega _{p} = 2\pi c/ \lambda_p$ represent small energy values $\frac{1}{2} \hbar \omega _{p}(L)$ in the contributions to the potential energy $U$, which therefore varies more slowly when $h$ increases above $100$ nm. 

In comparison,
using cylinders instead of cones (same hexagonal
lattice parameter and radius of $2.5$ nm) cannot provide efficient antireflection character to the surface since there is no gradient of the effective permittivity: the
reflection does not significantly decrease as the cylinder height increases (Fig. 4). As a result, the contact angle on this corrugated
surface saturates quickly with the cylinder height and stays close to the flat surface value (see inset table in Fig. 4), i.e. superhydrophobicity is never achieved.

\begin{figure}[t]
\centerline{\ \includegraphics[width=8 cm]{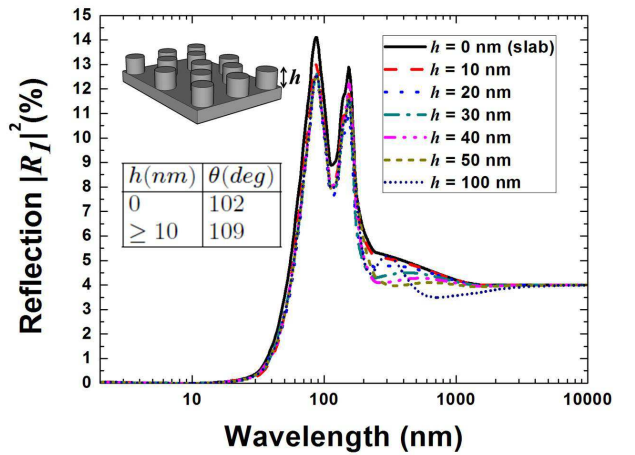}}
\caption{(Color online). Evolution of the reflectance with the height of
nanocylinders covering polyethylene slab. inset (table): evolution of the contact angle of
a water droplet on nanostructured polyethylene with the height \textit{h} of cylinders.}
\label{fig5}
\end{figure}

In summary, small nanoscopic corrugations dramatically affect the van der Waals interaction energy
and thus the wetting contact angle of molecular solid surfaces. This aspect was usually ignored in literature. The effect of nanostructuration on vacuum photon-modes should therefore be considered when wetting
phenomena are studied, because it is responsible for the interplay between superhydrophobicity and antireflection
property. Actually, in many cases, fractal structures appear to be quite efficient in achieving superhydrophobicity \cite{38,37}. Since those hierarchical structures involve, among others, nanoscopic corrugations (ranging from 10 nm to 100 nm) \cite{38,37}, their role in wetting phenomena deserves a reinterpretation in the light 
of the quantum vacuum photon-modes origin of the van der Waals force. 
Since our method relies on numerical solution of Eq. (1), it can also handle complex periodic corrugation geometries or non-flat meniscus by using more sophisticated numerical electromagnetic codes \cite{37b}. Finally, although the nanostructures referred in this study are quite small, many fabrication
techniques are available \cite{28,29,30,31,32,33,34,35,36,36b}. The fabrication of such surfaces could validate the theory presented in this letter. The crucial role of shallow corrugations (at the scale of 10 nm) on wettability was not noticed previously. Therefore, exploring corrugation dimensions where the effects here reported are dominant could pave the way for the design of novel superhydrophobic coatings. We believe the ab initio calculation of van der Waals force in nanostructures might contribute to elaborate more suitable models for describing more complex wettability phenomena such as contact angle hysteresis, Cassie state metastabilty and Cassie-Wenzel transition.

\textbf{Acknowledgments}

L.D. is supported by the Belgian Fund for Industrial and Agricultural
Research (FRIA). M.S. is supported by the Cleanoptic project (development of
super-hydrophobic anti-reflective coatings for solar glass panels/convention
No.1117317) of the Greenomat program of the Wallonia Region (Belgium). This
research used resources of the ``Plateforme Technologique de Calcul
Intensif'' (PTCI) (http://www.ptci.unamur.be) located at the University of
Namur, Belgium, which is supported by the F.R.S.-FNRS. The PTCI is member of
the ``Consortium des Equipements de Calcul Intensif (CECI)''
(http://www.ceci-hpc.be). The authors thank Philippe Lambin and François Fontaine for
critical reading of the manuscript.


\begin{thebibliography}{99}

\bibitem{1} D. Quéré, Physica A \textbf{313}, 32 (2002).

\bibitem{2} C. Yang, U. Tartaglino, and B. N. J. Persson, Phys. Rev. Lett. \textbf{97}, 116103 (2006).

\bibitem{2b} J.D. van der Waals, Ph.D. Thesis, Leiden University, 1873.

\bibitem{3} W. Fenzl, Ber. Bunsenges. Phys. Chem \textbf{98}, 389 (1994).

\bibitem{4} C. J. Drummond and D. Y. C. Chan, Langmuir \textbf{13}, 3890 (1997).

\bibitem{5} W. Fenzl, Europhys. Lett. \textbf{64}, 64 (2003).

\bibitem{5b}  A.B.D Cassie and S. Baxter, Trans. Faraday Soc., \textbf{40},
546 (1944).

\bibitem{5c}  C. Cuvelier, A. Segal, and A. A. van Steenhoven, \textit{Finite Element Methods and Navier-Stokes Equations}, Springer Science \& Business Media, 1986.

\bibitem{5cd} K. L. Mittal, \textit{Advances in Contact Angle, Wettability and Adhesion}, John Wiley \& Sons, 2013.

\bibitem{5cde} X. Yao, Y. Hu, A. Grinthal, T.-S. Wong, L. Mahadevan, and J. Aizenberg, Nat. Mater. \textbf{12}, 529 (2013).

\bibitem{5d}  J. De Coninck and T. D. Blake, Ann. Rev. Mater. Res. \textbf{38}, 1 (2008).

\bibitem{5e}  C. Yang, U. Tartaglino, and B. N. J. Persson, Phys. Rev. Lett. \textbf{97}, 116103 (2006).

\bibitem{6} H. B. G. Casimir, Proc. K. Ned. Akad. Wet. \textbf{51}, 793 (1948).

\bibitem{7} A. Lambrecht and V. N. Marachevsky, Phys. Rev. Lett. \textbf{101}, 160403 (2008).

\bibitem{8} M. T. H. Reid, A. W. Rodriguez, J. White, and S. G. Johnson, Phys. Rev. Lett. \textbf{103}, 040401 (2009).

\bibitem{9} P. S. Davids, F. Intravaia, F. S. S. Rosa, and D. A. R. Dalvit, Phys. Rev. A \textbf{82}, 062111 (2010).

\bibitem{10} F. S. S. Rosa, D. A. R. Dalvit, and P. W. Milonni, Phys. Rev. Lett. \textbf{100}, 183602 (2008).

\bibitem{11} E. M. Lifshitz, Sov. Phys. JETP \textbf{2}, 73 (1956).

\bibitem{12} A. Lambrecht, P. A. M. Neto, and S. Reynaud, New J. Phys. \textbf{8}, 243 (2006).

\bibitem{13} R. Messina and M. Antezza, Phys. Rev. A \textbf{84}, 042102 (2011).

\bibitem{14} J. Mahanty, B.W. Ninham, \textit{Dispersion Forces}, Academic Press London, 1976.

\bibitem{14c} J. Israelachvili, \textit{Intermolecular and surface forces}, Academic Press Elsevier, 2011.

\bibitem{14d} D.B Hough, and L.R White, Adv. Colloid Interfac.\textbf{14}, 3 (1980.

\bibitem{14b} Hamaker, H. C. Physica \textbf{4}, 1058 (1937).

\bibitem{14e} Strickly, the $L_0$ value should depend on the surface roughness for the following reason. Whereas the short-range Pauli repulsive term in the Lennard-Jones-like interfacial potential remains unchanged while increasing the cones height since it is determined by the topmost atomic layer, the van der Waals attractive term, on the contrary, decreases (Fig. 2b). As a result, the potential minimum shifts towards longer distances, i.e. $L_0$ increases. The cumulative effect, i.e. increase of $L_0$ and decrease of $A_h$ (Fig. 3a) while increasing $h$, reinforces the hydrophobic character according to Eq. (3).

\bibitem{15}  H.-J. Butt, K. Graf, and M. Kappl, \textit{Physics and Chemistry of
Interfaces}, Wiley-VCH Verlag GmbH \& Co. KGaA, 2003.

\bibitem{17}  D. G. Stavenga, S. Foletti, G. Palasantzas, and K. Arikawa, Proc. Biol. Sci. \textbf{273}, 661 (2006).

\bibitem{18}  L. Dellieu, M. Sarrazin, P. Simonis, O. Deparis, and J. P. Vigneron, J. Appl. Phys. \textbf{116}, 024701 (2014).

\bibitem{23}  A. Deinega and S. John, Opt. Lett. \textbf{37}, 112 (2012).

\bibitem{24}  J. Ashok, P.L.H. Varaprasad, and J.R. Birch, in \textit{Handbook of Optical Constants
of Solids II}, edited by E.D. Palik (Academic Press Elsevier, New York, 1991).

\bibitem{25}  V. A. Parsegian and G. H. Weiss, J. Colloid Interf. Sci. \textbf{81}, 285 (1981).

\bibitem{25b} S. Mouchet, O. Deparis, and J.-P. Vigneron, Proc. SPIE 8424, 842425 (2012).

\bibitem{26}  C. Chen, P. Berini, D. Feng, S. Tanev, and V. Tzolov, Opt. Express\textbf{ 7}, 260 (2000).

\bibitem{27} N. De Geyter, R. Morent, and C. Leys, Surf. Interface Anal. \textbf{40}, 608 (2008).

\bibitem{photonics} B. E. A. Saleh and M. C. Teich, \textit{Fundamentals of Photonics}, Wiley-Interscience, 2 edition, 2007.

\bibitem{38} S. Shibuichi, T. Onda, N. Satoh, and K. Tsujii, J. Phys. Chem. \textbf{100}, 19512 (1996).

\bibitem{37} B. Bhushan, Y. C. Jung, and K. Koch, Philos. Trans. A. Math. Phys. Eng. Sci. \textbf{367}, 1631 (2009).

\bibitem{37b} M. Sarrazin, J.-P. Vigneron, and J.-M. Vigoureux, Phys. Rev. B \textbf{67}, 085415 (2003).

\bibitem{34} X. Lu, C. Zhang, and Y. Han, Macromol. Rapid Commun. \textbf{25}, 1606 (2004).

\bibitem{36} C.-T. Hsieh, J.-M. Chen, R.-R. Kuo, T.-S. Lin, and C.-F. Wu, Appl. Surf. Sci \textbf{240}, 318 (2005).

\bibitem{28}  A. Checco, B. M. Ocko, A. Rahman, C. T. Black, M. Tasinkevych,
A. Giacomello, and S. Dietrich, Phys. Rev. Lett. \textbf{112}, 216101 (2014)  

\bibitem{29}  H. Tani, H. Takahashi, S. Koganezawa, and N. Tagawa, Tribol. Lett. \textbf{54}, 221 (2014).

\bibitem{30}  Y. P. Li, Z. C. Zhang, W. Shi, and M. K. Lei, Surf. Coat. Tech. \textbf{259}, 77 (2014).

\bibitem{31} J. Fresnais, J. P. Chapel, and F. Poncin-Epaillard, Surf. Coat. Tech. \textbf{200}, 5296 (2006).

\bibitem{32} E. Burkarter, C. K. Saul, F. Thomazi, N. C. Cruz, L. S. Roman, and W. H. Schreiner,Surf. Coat. Tech.  \textbf{202}, 194 (2007).

\bibitem{33} J. Zimmermann, M. Rabe, G. R. J. Artus, and S. Seeger, Soft Matter \textbf{4}, 450 (2008).

\bibitem{35}  W. Du and J. Di, Surf. Coat. Tech. \textbf{201}, 5498 (2007).

\bibitem{36b} E. Martines, K. Seunarine, H. Morgan, N. Gadegaard, C. D. W. Wilkinson, and M. O. Riehle,  Nano letters, \textbf{5}, 2097 (2005).


\end{thebibliography}
\end{document}